\documentclass[twocolumn]{revtex4}
\usepackage{jpc}
\usepackage{graphicx}

\begin{document}

\title{{\em Ab initio} theory of Cr$_2$O$_3$ surface chemistry in solution}

\author{S. Petrosyan}
\affiliation{Department of Physics, Cornell University, Ithaca, NY 14853}

\author{A.A. Rigos}
\affiliation{Department of Chemistry, Merrimack College, North Andover, MA}

\author{T.A. Arias}
\affiliation{Department of Physics, Cornell University, Ithaca, NY 14853}

\begin{abstract}
Using a new form of density functional theory for the {\em ab initio}
description of electronic systems in contact with a dielectric
environment, we present the first detailed study of the impact of a
solvent on the surface chemistry of Cr$_2$O$_3$, the passivating layer
of stainless steel alloys.  Compared to vacuum, we predict that the
presence of water has little impact on the adsorption of chloride ions
to the oxygen-terminated surface but a dramatic effect on the binding
of hydrogen to that surface.  These results indicate that the
dielectric screening properties of water are important to the
passivating effects of the oxygen-terminated surface.

\end{abstract}

\maketitle

\section{Introduction}

{\em Ab initio} calculations have shed light on many physico-chemical
questions, including chemical reactions in solution and chemical
reactions at surfaces \cite{Radeke, Greeley, Gross}.  Although
computational chemistry is now able to provide not only qualitative
but also quantitative insights into surface chemistry \cite{Greeley},
so far these studies have been limited to reactions in a vacuum even
though experimentally these reactions often occur in a solvent
environment.

Each year, corrosion costs the United States \$276 billion
\cite{DeGaspari}, approximately 3.1\% of the gross domestic product,
and many approaches have been used to understand and model this
complex process \cite{Ryan, Stampfl, Erlebacher}. High performance
stainless steel alloys contain chromium to result in the formation of
a Cr$_2$O$_3$ passivating surface layer, which provides corrosion
resistance.  Even with this layer, such alloys are susceptible to
breakdown in acidic, chlorine-containing aqueous
environments\cite{Jones,Lats}, the study of which demands simultaneous
treatment of surfaces, reactants and the aqueous dielectric
environment.  Direct experiments are difficult, and although {\em ab
initio} calculations have a history of answering many such questions,
to date they have not been able to address the effect of the solvent
on surface reactions.  Alavi {\em et al.}\cite{alavi}, for instance,
have studied the adsorption of HCl on single-crystal alpha-Al$_2$O$_3$
(0001) surfaces and calculated adsorption energies as a function of
surface coverage within density-functional theory, but all of their
results are obtained in a vacuum environment.

Previous studies of Cr$_2$O$_3$ have been limited to pure surfaces in
vacuum \cite{Cline}, an unrealistic environment for the study of
corrosion.  These studies indicated that the highest occupied and
lowest unoccupied molecular orbitals are localized on the chromium
ions, suggesting that the oxygen-terminated surface could provide a
stable barrier against acidic chlorinated environments {\em if} the
surface oxygen layer could be prevented from reacting with species in
the solution.  The work below puts forth evidence to support the novel
hypothesis that the dielectric screening effects associated with an
aqueous environment actually prevent the formation of bonds with
aqueous species such as protons, thereby rendering the
oxygen-terminated surface virtually non-reactive.

Models for calculating solvation energies from continuum dielectric
theory\cite{Tannor} have been applied to single molecules or activated
complexes but not to molecules adsorbed on surfaces, perhaps because
such methods are generally applied to molecules.  Here, we introduce
the first approach to {\em ab initio} calculations in a dielectric
environment which sits on a firm theoretical foundation.  Below, we
show that this new approach, which in a simple approximation is
related to that recently introduced by Fattebert and Gygi\cite{gygi1},
gives results in good agreement with currently accepted quantum
chemical methods and is well-suited to surfaces.  Finally, we apply
the approach to carry out the first {\em ab initio} study of the
reactivity of hydrogen and chlorine on an oxygen-terminated
Cr$_2$O$_3$(0001) surface in contact with a solution.

\section{Theoretical approach}

Because of the large cells required in this study (well over one
hundred atoms), the only practicable {\em ab initio} approach suited
to describe the electrons of the surface and the reactants is
density-functional theory\cite{Cline}.  The periodicity of the surface
makes periodic boundary conditions and thus the plane-wave
pseudopotential method \cite{rmp} the most natural choice.  The size
of the surface cell and time-scales needed for proper thermodynamic
averaging make a direct molecular dynamics treatment of the aqueous
environment infeasible, thus raising the question of how to treat the
solvent.  This work takes the novel approach of exploiting the
existence of so-called ``classical'' density-functional
theories\cite{Ashcroft}, which can treat water rigorously in terms of
a simple thermodynamically averaged molecular density and an
approximate functional\cite{Sun}.  We then introduce here for the
first time the concept of a {\em joint density-functional theory}
(JDFT) between the electrons in the surface and the molecules
comprising the solvent.

In a companion to this paper\cite{theory1.0}, we prove that the total
thermodynamic free energy $A$ of an electronic system (solute) with
nuclei of charges $Z_I$ at locations $R_I$ in equilibrium with a
closed-shell liquid molecular environment (solvent) may be determined
from the variational principle
\begin{eqnarray} 
A&=&\min_{n(r),N(r)}\left( A_{KS}[n(r),\{Z_I,R_I\}]
\label{eq:JDFT1}  \right. \\
&& \mbox{\ \ } \left. + A_{lq}[N(r)]+U[n(r),N(r),\{Z_I,R_I\}]\right), \nonumber
\end{eqnarray}
where, at the minimum, $n(r)$ is the thermo- and quantum- mechanically
averaged density of the electrons of the solute and $N(r)$ is the
likewise averaged molecular density of the solvent.
$A_{KS}[n(r),\{Z_I,R_I\}]$, $A_{lq}[N(r)]$ and
$U[n(r),N(r),\{Z_I,R_I\}]$ appearing above are, respectively, the
standard Kohn-Sham electron-density functional of the solute when in
isolation, the classical density-functional for the molecular solvent
when in isolation, and a new functional describing the coupling
between the systems.  The new functional $U[n(r),N(r),\{Z_I,R_I\}]$ is
universal in the sense that it depends only upon the nature of the
solvent and is independent of the nature of the solute.  For
completeness, we note that each of the three aforementioned
functionals is implicitly also a function of the temperature $T$.

In the present work, we make the further simplification of performing the
minimization over $N(r)$ in (\ref{eq:JDFT1}), resulting in the
variational principle
\begin{eqnarray} 
A&=&\min_{n(r)}\left(A_{KS}[n(r),\{Z_I,R_I\}]\right. \label{eq:JDFT2} \\
&& \mbox{\ \ \ }\left.+W[n(r),\{Z_I,R_I\}]\right), \nonumber
\end{eqnarray}
where
\begin{eqnarray*}
W[n(r),\{Z_I,R_I\}] & \equiv & 
\min_{N(r)}\left(A_{lq}[N(r)] \right.\\
&& \mbox{\ \ \ } \left. +U[n(r),N(r),\{Z_I,R_I\}]\right)
\end{eqnarray*}
is a universal functional dependent solely upon the identity of the
environment (and, implicitly, the temperature).  Note that, in
principle, the theory at this stage is exact.  Below we outline the
approximations which we introduce because the exact form of the
functionals $A_{KS}[n(r),\{Z_I,R_I\}]$ and $W[n(r),\{Z_I,R_I\}]$ are
unknown.

\section{Computational Details}

For treatment of the electrons in the chromium-oxide surface through
the functional $A_{KS}[n(r),\{Z_I,R_I\}]$, we apply the standard local
spin-density approximation (LSDA)\cite{Perdew}. The calculations
themselves employ the total-energy plane-wave density-functional
pseudopotential approach~\cite{rmp} with potentials of the
Kleinman-Bylander form~\cite{KB} with $p$ and $d$ non-local
corrections at a cutoff of 40~hartrees.  Supercells with periodic
boundary conditions in all three dimensions represent the surfaces of
isolated oxygen-terminated (0001)-oriented slabs of Cr$_2$O$_3$ of
thickness 13~\AA\ separated by 7.8~\AA\ of either vacuum or solvent.
The in-plane boundary conditions suffice to describe $2 \times 2$
reconstructions and consist of four times the unit from \cite{Cline}
so as to allow sufficient isolation of solvent volumes excluded by
chlorine adsorbed on the surface.  The supercell contains a total of
forty chromium and seventy-two oxygen atoms with chlorine or hydrogen
added to the two surfaces in inversion-symmetric pairs, one on each
side of the slab.  Finally, we use a single k-point to sample the
Brillioun zone of the surface slab.  Ref.~\cite{Cline} establishes
that this choice of functional, pseudopotential, plane-wave cutoff,
sampling density in the Brillioun zone and supercell gives a good
description of the bulk and surface of Cr$_2$O$_3$.  As in the
aforementioned work, we employ the analytically continued functional
approach~\cite{rmp,ACprl} to minimize the Kohn-Sham energy with
respect to the electronic degrees of freedom.  Below, we relax all
ionic configurations until the total energy is within 0.027~eV of the
minimum and the maximum force in any direction is less than
0.3~eV/\AA.

For the environment functional $W[n(r),\{Z_I,R_I\}]$ in
(\ref{eq:JDFT2}), we take the interaction of the electron and nuclear
charges of the system under study with a dielectric environment in which
the dielectric constant is local in space and has a value
dependent only upon the electron density at each point,
$\epsilon(r)\equiv \epsilon(n(r))$.  In practice, we compute the total
free energy of the system in contact with the environment by finding
the stationary point with respect to both the electrons and the mean
electrostatic field $\phi(r)$ of the functional
\begin{eqnarray}
A & = &
A_{TXC}[n_\uparrow(r),n_\downarrow(r)] + \Delta V_{ps}[n_\uparrow(r),n_\downarrow(r)] \label{eq:aJDFT} \\
&& + \int d^3r \, \left\{\phi(r) \left(n_{tot}(r)-\sum_I Z_I
\delta^{(3)}(r-R_I)\right) \right. \nonumber \\
&&  \left. - \frac{\epsilon(n(r))}{8\pi} \left| \nabla \phi(r)
\right|^2\right\} \nonumber
\end{eqnarray}
where $n_\uparrow(r)$, $n_\downarrow(r)$ and $n_{tot}(r)$,
respectively, are the \mbox{up-,} down- and total electron densities,
$A_{TXC}[n_\uparrow(r),n_\downarrow(r)]$ is the Kohn-Sham
single-particle kinetic plus exchange correlation energy within the
local spin-density approximation, $\Delta V_{ps}$ is the difference in
the total pseudopotential energy from that expected from pure Coulomb
interactions with point ions of valence charges $Z_I$ at locations
$R_I$, and $\delta^{(3)}(r)$ is the three-dimensional Dirac-$\delta$
function.  Note that, although we do work directly with the Kohn-Sham
orbitals, for brevity we have written the above in terms of the
electron densities.  We also note that at this level of approximation
our joint density functional theory takes the same form as the
approach of \cite{gygi1}, which introduced the form as a computational
device without formal justification or proof.

\begin{figure} \centering
\includegraphics[width=3.5in]{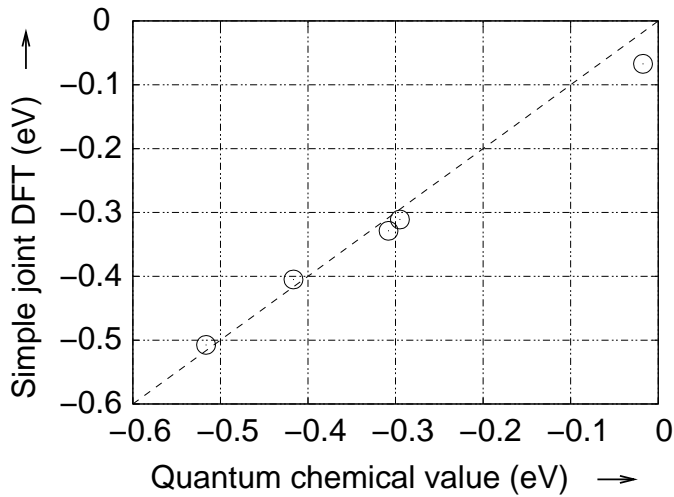}
\caption{Comparison of predictions of a simple joint density-functional
theory (vertical axis) with established quantum chemical values
(horizontal axis) for acetamide, acetic acid, methanol, ethanol and
methane (from left to right) from \cite{Marten}.}
\label{fig:lda}
\end{figure}

For the local dielectric function $\epsilon(n(r))$, we choose a
specific form which varies smoothly from the dielectric constant of
the bulk solvent $\epsilon_b$ when the electron density $n(r)$ is less
than a critical value $n_c$ indicative of the interior of the solvent
to the dielectric constant of vacuum $\epsilon=1$ when $n(r)>n_c$.
Specifically, we take
\begin{equation}
\epsilon(n) = 1+\frac{\epsilon_b-1}{2}
\mbox{\,erfc}\left(\frac{\ln(n/n_c)}{\sqrt{2}\sigma}\right), \label{eq:lda}
\end{equation}
where the parameter $\sigma$, to which the results are not very
sensitive, controls the width of the transition from bulk to vacuum
behavior.  To determine the parameters $\sigma$ and $n_c$ for this
simple model, we fit computed electrostatic solvation energies in
aqueous solution ($\epsilon_b=80$) to the accepted values from the
quantum-chemical literature for methane, ethanol, methanol, acetic
acid and acetamide\cite{Marten}.  Figure \ref{fig:lda} summarizes the
quality of this comparison for our final choice of fit parameters,
$\sigma=0.6$ and $n_c = 4.73 \times 10^{-3}$~\AA$^{-3}$.

\section{Results and discussion}

\subsection{Pristine surface}

\begin{figure} \centering
\includegraphics[width=3.4in]{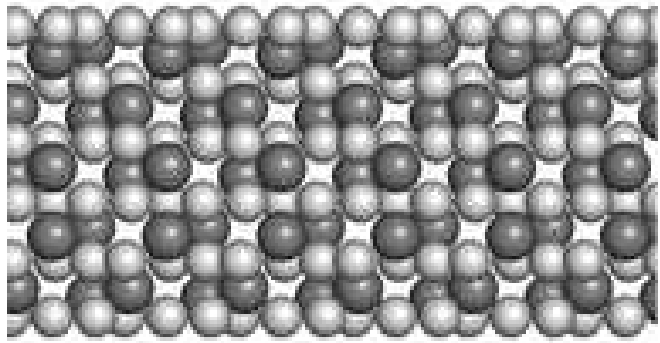}
  \caption{Relaxed structure of pristine surface slab, side view
    ($[0001]$ direction runs vertically up the page): oxygen (light
    grey spheres), chromium (dark grey spheres).}
  \label{fig:bulkras}
\end{figure}

Ref.~\cite{Cline} reviews in detail the structure of bulk Cr$_2$O$_3$
and the relaxation of its pristine (0001) oxygen-terminated surface.
Figure~\ref{fig:bulkras} shows the relaxed structure of our supercell
surface slab.  The bulk structure consists of alternating planar
layers of oxygen atoms separated by bilayers of chromium.  As found in
\cite{Cline}, the primary relaxation associated with forming the
surface is for the oxygen-terminated surface layers to move inward
toward the bulk crystal with slight in-plane displacements.

\begin{figure} \centering
  \includegraphics[width=3.5in]{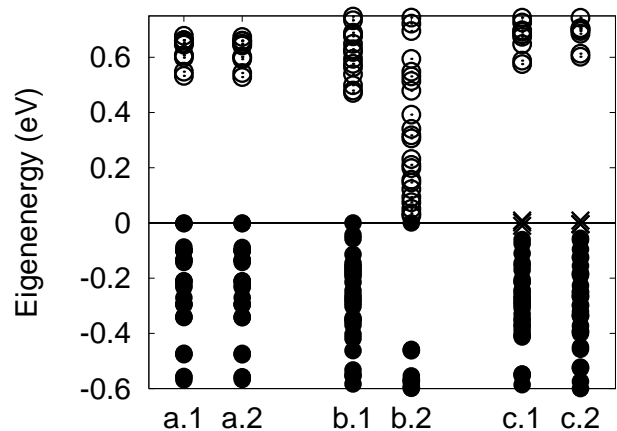}
  \caption{Near-gap energy levels for (a) pristine, (b) hydrogenated
    and (c) chlorinated system in (1) vacuum and (2) dielectric:
    filled levels (solid circles), empty levels (open circles),
    partially filled levels (crosses).}
  \label{fig:elevels}
\end{figure}

Figure~\ref{fig:elevels}a.1 shows the filled and empty energy levels
from our supercell calculation of the (0001) oxygen-terminated surface
of Cr$_2$O$_3$ in a vacuum environment.  Following standard practice,
we choose the zero of energy to be the Fermi level, the energy below
which states are fully occupied and above which they are empty.  The
states at the zero line in the figure are thus the highest occupied
molecular orbitals (HOMOs) and the first states above the line are the
lowest unoccupied molecular orbitals (LUMOs).  The figure shows the
levels of the pristine surface to be fully filled up to a gap of about
0.5~eV separating the HOMOs and LUMOs.

\begin{figure}
  \centering
  \includegraphics[width=3.5in]{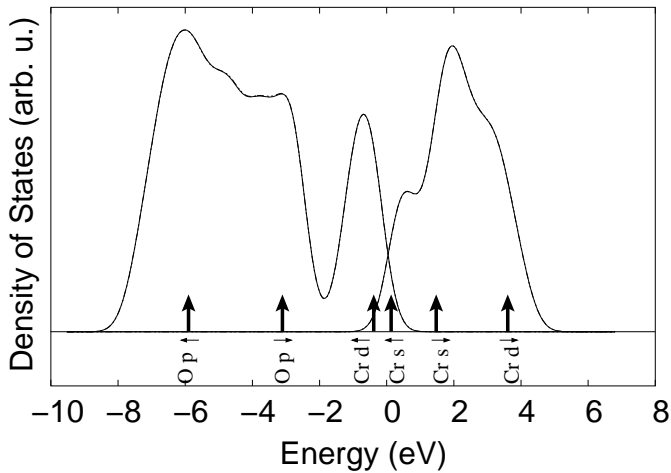}
  \caption{Total density of states of pristine supercell slab in vacuum (solid
  curve) and in solution (virtually indistinguishable dashed curve):
  occupied states (curve on left), unoccupied states (curve on right),
  LSDA atomic eigenvalues (vertical arrows) labeled according to
  alignment ($\uparrow$) or anti-alignment ($\downarrow$) with
  direction of net atomic spin.} \label{fig:tdos}
\end{figure}

To provide a more global view, Figure~\ref{fig:tdos} presents the
density of states, the number of levels from
Figure~\ref{fig:elevels}a.1 per unit energy as a function of energy,
computed using a Gaussian broadening of width $\sigma$=0.41~eV. To
underscore the distinction between occupied and unoccupied states, the
figure gives a separate curve for each.  Finally, as a guide to
identification of the bands, the figure also contains markers for the
LSDA atomic eigenvalues of oxygen and chromium, which have been
shifted uniformly upward by 4.4~eV to approximately counteract the
shifting of the Fermi level of the supercell states to zero energy.

\begin{figure}
\centering 
\includegraphics[height=3.5in]{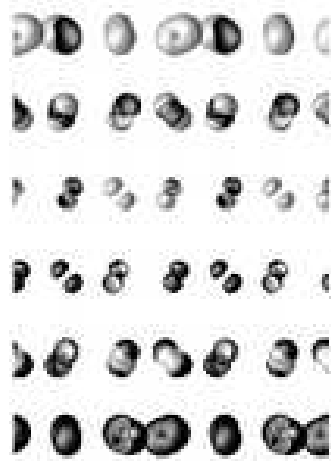}
\caption{Contour levels of sum of probability densities from the highest
energy shoulder of the oxygen $2p$ band, side view ($[0001]$ direction
runs vertically up the page): up spin (black surface), down spin (white surface).}
\label{fig:Osmap} \end{figure}

\begin{figure}
  \centering	
  \includegraphics[height=3.5in]{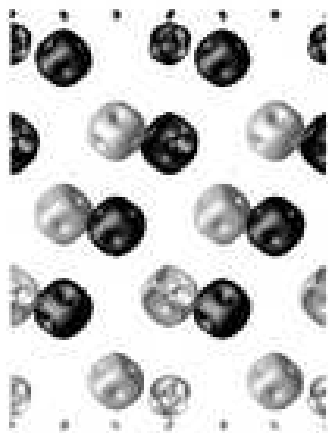}
  \caption{Contour levels of sum of probability densities from the
  chromium $sd\uparrow$ band, side view ($[0001]$ direction
runs vertically up the page): up spin (black surface), down spin (white surface).} \label{fig:Crmap}
  \end{figure}

\begin{figure}\centering
\includegraphics[height=3.5in]{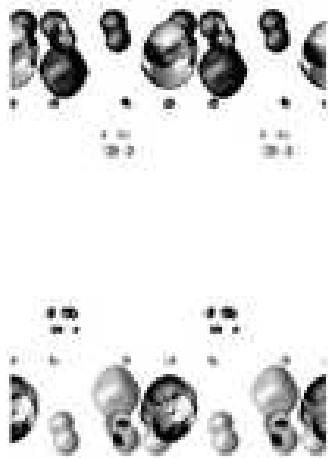}
  \caption{Contour level of sum of probability densities from the
lowest energy shoulder of the chromium $sd\downarrow$ spin band, side
view ($[0001]$ direction runs vertically up the page): up spin (black surface), down spin (white surface).}
\label{fig:Esmap} \end{figure}

Three features in the supercell density of states play important roles
in the chemistry of this surface.  First, the highest occupied {\em
oxygen} orbitals appear as the shoulder (from $\sim$-4~eV to
$\sim$-2~eV) of the oxygen $2p$ band, which consists of ``minority''
spin electrons, electrons with spin opposite to the net atomic spin.
Figure~\ref{fig:Osmap}a presents the sum of the probabilities
associated with states in this range and clearly indicates this
shoulder to be a surface oxygen band.  The localization of different
spin types to either surface corresponds to the fact that the oxygen
atoms on each surface have opposite net spin direction.  Next, the
highest occupied molecular orbitals (HOMOs) overall appear as an
$sd$-hybrid chromium band (from $\sim$-2~eV to $\sim$0~eV) consisting
of electrons of ``majority'' spin, spin aligned with the net atomic
spin.  Figure~\ref{fig:Crmap} shows these states to be {\em bulk}
chromium states, with atoms alternating in majority spin direction
corresponding to the anti-ferromagnetic nature of the bulk material.
Finally, the LUMOs of the system appear as the low energy majority
spin shoulder (from $\sim$0~H to $\sim$1~eV) of an empty chromium
band.  Figure~\ref{fig:Esmap} shows this shoulder to consist of
surface states of primarily chromium character protected under the
outer oxygen layer.  This character of the HOMOs and LUMOs suggests
that any oxygen missing from the outer surface would expose a reactive
chromium layer underneath.

Upon repeating the pristine surface calculation in the presence of a
dielectric environment, we find virtually no change.  There is no
change in Figure~\ref{fig:bulkras} and, although some very small
changes are evident in going from Figure~\ref{fig:elevels}a.1 to
Figure~\ref{fig:elevels}a.2, the global picture of the density of
states in Figure~\ref{fig:tdos} is visually indistinguishable for
vacuum (solid curve) and dielectric (dashed curve).  Finally,
inspection of density maps corresponding to
Figures~\ref{fig:Osmap}--\ref{fig:Esmap} again shows no noticeable
changes.

\subsection{Interaction with hydrogen}

\begin{figure}
  \centering
\parbox{1.5in}{\includegraphics[width=1.5in]{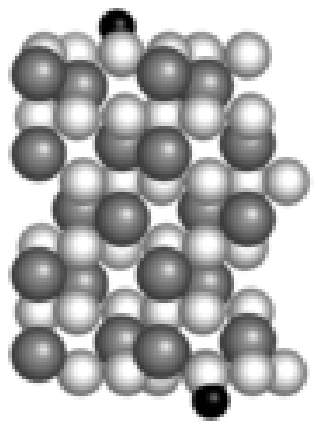}\\(a)}
\parbox{1.5in}{\includegraphics[width=1.5in]{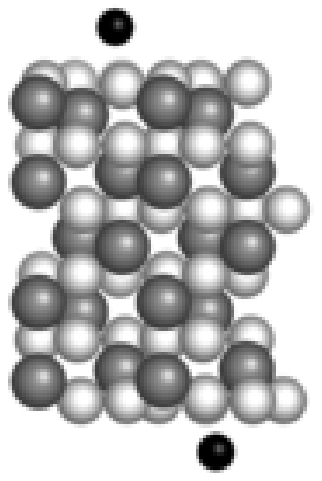}\\(b)}
\caption{Relaxed structure of surface slab with adsorbed hydrogen in
  (a) vacuum and (b) dielectric: oxygen (light grey spheres), chromium
  (dark grey spheres), hydrogen (black spheres).  Same view as
  Figure~\ref{fig:bulkras} but showing atoms from a single
  supercell.}
\label{fig:hvacrasside}
\end{figure}

\begin{figure}
  \centering
  \includegraphics[width=3.5in]{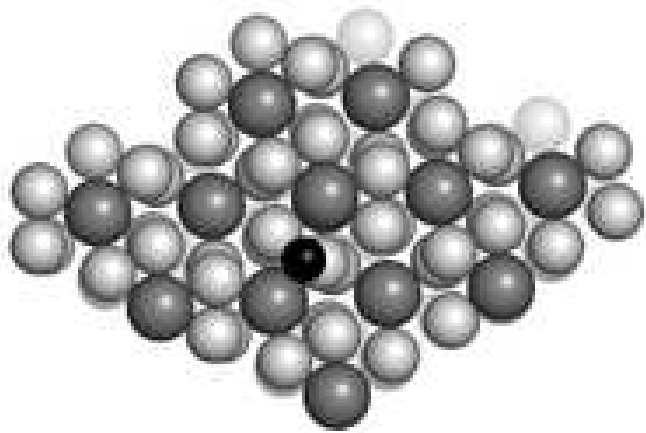}
  \caption{Relaxed structure of single supercell with adsorbed hydrogen in
    vacuum, top view ($[0001]$ direction normal to page): oxygen 
    (light grey spheres), chromium (dark grey spheres), hydrogen
    (black sphere).}
  \label{fig:hvacrastop}
\end{figure}

Anticipating bonding with oxygen, we initially placed a hydrogen atom
in vacuum directly on top of one of the surface oxygen atoms, all of
which are equivalent by symmetry.  Figure~\ref{fig:hvacrasside}a shows
that, upon relaxation, the hydrogen atom cants away from the surface
perpendicular while appearing to form a bond with the underlying
oxygen atom: the final relaxed O-H distance is 0.95~\AA, quite close
to the experimental O-H separation in H$_2$O (0.96~\AA).
Figure~\ref{fig:hvacrastop} shows that the canting of the hydrogen
atom is in the same direction as one would expect for the $2p$ orbital
of the associated oxygen atom given its in-plane displacement.
Finally, Figure~\ref{fig:Hdensvac}a, which shows the total electron
density associated with the chemisorbed H, confirms the presence of
the bond as a small protrusion in the density near the hydrogen atom.

\begin{figure}
  \centering 

\parbox{1.5in}{\includegraphics[width=1.4in]{figures/hyd/vac/n.epsf}\\(a)}
\parbox{1.5in}{\includegraphics[width=1.4in]{figures/hyd/diel/n.epsf}\\(a)}
  \caption{Contour level of 0.68~$e^-$/\AA$^{3}$ total electron
  density for hydrogen atom adsorbed on oxygen-terminated $(0001)$
  surface in (a) vacuum and (b) dielectric, side view ($[0001]$ direction
runs vertically up the page).}
  \label{fig:Hdensvac}
\end{figure}

\begin{figure}
\centering
\includegraphics[width=3.5in]{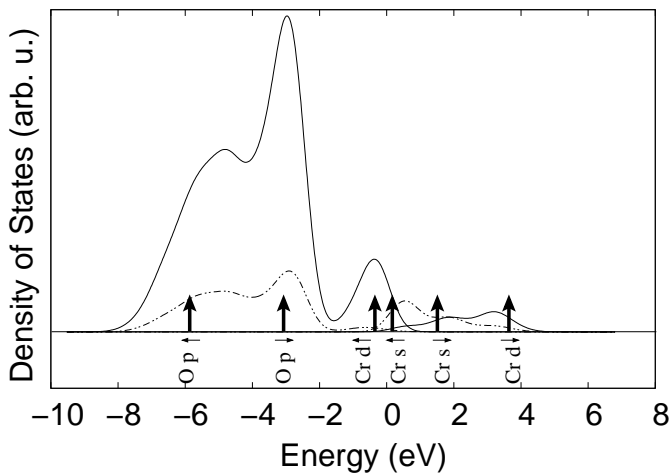}
\caption{Local density of states within
0.69~\AA\ of the proton for hydrogen interacting with oxygen-terminated
$(0001)$ surface in
vacuum (solid curve) and solution (dashed curve).  Same conventions as Figure~\ref{fig:tdos}.}  \label{fig:hdos}
\end{figure}

\begin{figure}
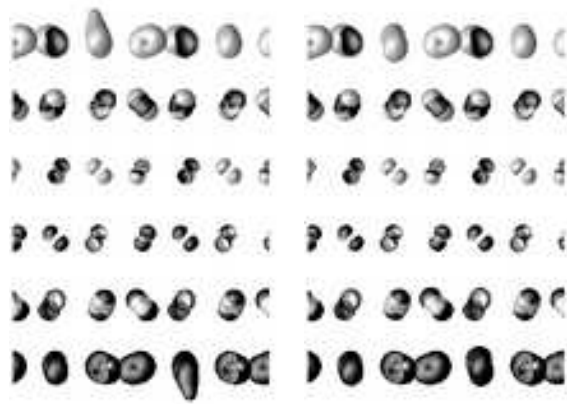

  \centering 
\parbox{1.5in}{\includegraphics[width=1.4in]{figures/hyd/vac/homo_414_447.epsf}\\(a)}
\parbox{1.5in}{\includegraphics[width=1.4in]{figures/hyd/diel/homo_414_447.epsf}\\(b)}
\caption{Contour level of sum of probability densities associated with
oxygen $2p$ surface band in (a) vacuum and (b) dielectric: up spin (black surface), down spin (white surface).}
\label{fig:OsmapH} \end{figure}

\begin{figure}
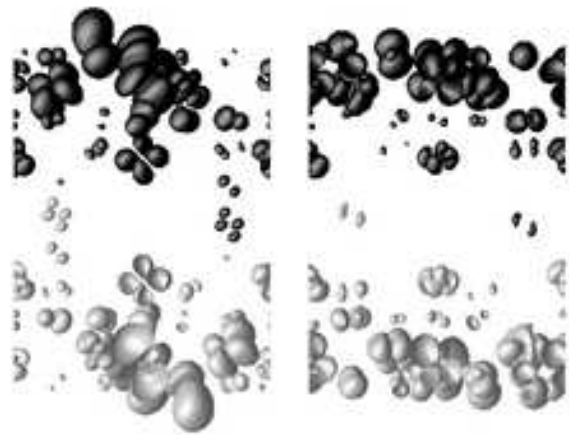

  \centering 
\parbox{1.5in}{\includegraphics[width=1.4in]{figures/hyd/vac/496.epsf}\\(a)}
\parbox{1.5in}{\includegraphics[width=1.4in]{figures/hyd/diel/496.epsf}\\(b)}
  \caption{Contour level of HOMO of hydrogen interacting with the
  oxygen-terminated $(0001)$ surface in (a) vacuum and (b) dielectric,
  side view ($[0001]$ direction runs vertically up the page): up spin (black surface), down spin (white surface).}
  \label{fig:hHOMO} \end{figure}

Figure~\ref{fig:elevels}b.1 shows the filled and empty energy levels
of the hydrogenated surface in a vacuum environment.  As with the
pristine surface (Figure~\ref{fig:elevels}a), the energy levels are
fully filled up to a HOMO-LUMO gap, consistent with the
observed bonding.  To better resolve the bond associated with the
chemisorbed hydrogen, Figure~\ref{fig:hdos} presents the {\em local}
density of states in the vicinity of the hydrogen atom, which we
compute in the same way as the total density of states of
Figure~\ref{fig:tdos} but by now weighing each level with the
probability of an electron in the level being within 0.69~\AA\ of the
proton.  The local density of states shows that the hydrogen atom
interacts mostly with the surface oxygen $2p$ band.  A plot of the
total density associated with this surface band,
Figure~\ref{fig:OsmapH}a, confirms that it contains most of the
density protrusion associated with the O-H bond.  Finally,
Figure~\ref{fig:hHOMO}a shows that the HOMO of the hydrogenated
surface, while maintaining significant bulk chromium character,
indeed localizes near the hydrogen atom.

To determine the final atomic configuration in the presence of a
solvent, we begin with the positions from
Figure~\ref{fig:hvacrasside}a and relax the atomic coordinates within
the approximate joint density functional (\ref{eq:aJDFT},\ref{eq:lda})
until the maximum force on any atom is less than 0.3 eV/\AA.
Figure~\ref{fig:hvacrasside}b, which displays the resulting
configuration, shows that the presence of the dielectric has a
dramatic effect on the hydrogen.  The nearest oxygen-hydrogen distance
has increased to 2.3~\AA, clearly breaking the O-H bond and returning
the hydrogen atom to the solution.  Consistent with this picture, the
largest residual force remains on the hydrogen atom in the direction
tending to push it yet further from the surface.  Lack of any
indication of the presence of the hydrogen atom in the resulting total
charge density, Figure~\ref{fig:Hdensvac}b, indicates that the atom
enters the solution as an ion.

Figure~\ref{fig:elevels}b.2 shows that upon the removal of the proton
from the surface, the excess electron from the O-H bond appears simply
as a donated conduction electron just above the energy gap.
Consistent with this donation, the dashed curve in
Figure~\ref{fig:hdos} shows a dramatic reduction in the local density
of states near the proton, and the HOMO in Figure~\ref{fig:hHOMO}b now
shows more of the surface chromium character of the original LUMO
band from Figure~\ref{fig:Esmap}.

\subsection{Interaction with chlorine}

\begin{figure}
  \centering
  \includegraphics[width=3.5in]{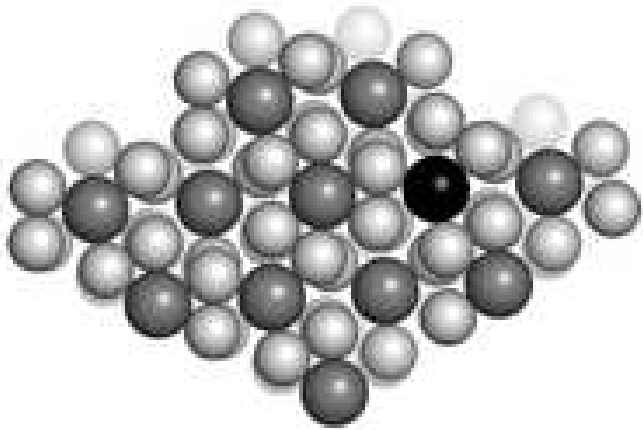}
  \caption{Relaxed structure of surface interacting with chlorine in
    vacuum: oxygen (light grey spheres), chromium (dark grey spheres),
    chlorine (black sphere): top view ($[0001]$ direction perpendicular
    to page).}
  \label{fig:clvacrastop}
\end{figure}

\begin{figure}
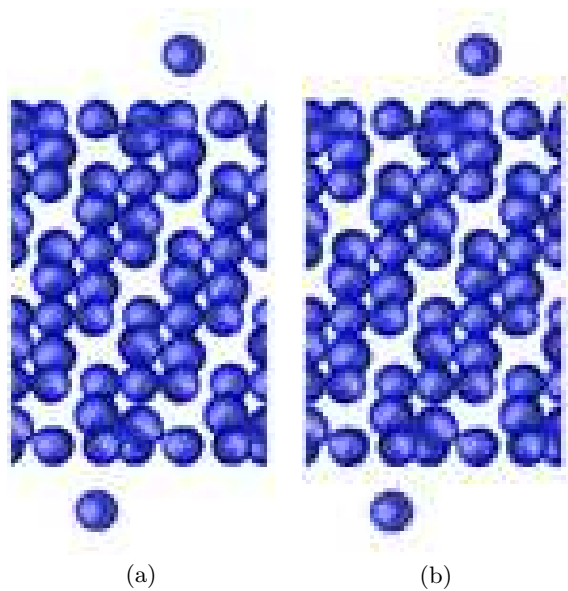

  \centering 
\parbox{1.5in}{\includegraphics[width=1.4in]{figures/cl/vac/n.epsf}\\(a)}
\parbox{1.5in}{\includegraphics[width=1.4in]{figures/cl/diel/n.epsf}\\(b)}
  \caption{Contour level of 0.68~$e^-$/\AA$^{3}$ total electron density for
  chlorine atom adsorbed on oxygen-terminated $(0001)$ surface in (a) vacuum and
(b) dielectric, side view ($[0001]$ direction
runs vertically up the page).}
  \label{fig:Cldensvac}
\end{figure}

Anticipating the possibilities of ionic bonding for chlorine, we
initially placed (in the 1$\times$1 surface supercell) a chlorine atom in
vacuum directly above each of the two distinct types of Cr$^{3+}$ site
from the outermost chromium bilayer and found the site above the
innermost of the two component layers to be favored by 0.4~eV.
Figure~\ref{fig:clvacrastop} shows a top ($[0001]$) view of the
relaxed configuration for this site when computed within the
4$\times$4 supercell.  We find relatively little relaxation from the
clean surface structure (movement of all surface atoms is less than
0.01~\AA) with the chlorine ion settling upon relaxation to a position
with a chlorine-oxygen separation of 2.6~\AA, only 20\% smaller than
the sum of the nominal ionic radii, 3.1~\AA.
Figure~\ref{fig:Cldensvac}a shows a side view of the total electron
density for a surface with adsorbed Cl in vacuum, illustrating the
physisorbed nature of the interaction.

\begin{figure}
  \centering
  \includegraphics[width=3.5in]{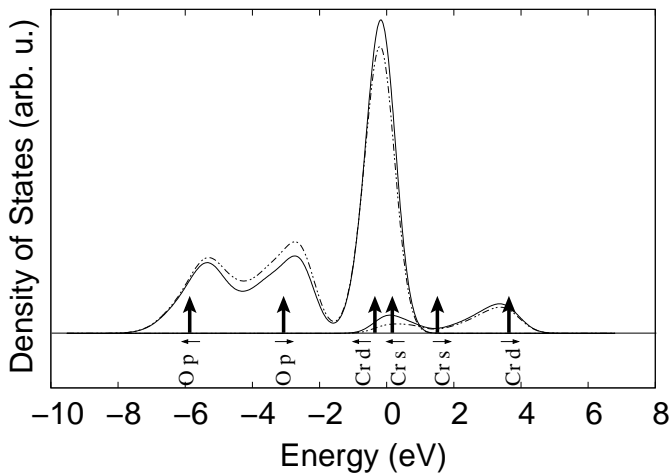}
  \caption{Local density of states within 1.6~\AA\ of the
   chlorine nucleus for chlorine interacting with oxygen-terminated
   $(0001$) surface in vacuum (solid curve) and solution (dashed
   curve).  Same conventions as Figure~\ref{fig:tdos}.}
  \label{fig:Cldos}
\end{figure}

\begin{figure}
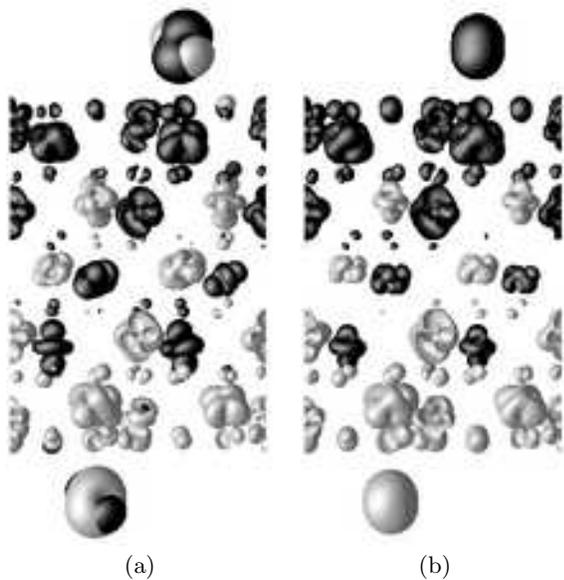

  \centering
\parbox{1.5in}{\includegraphics[width=1.4in]{figures/cl/vac/lumo_501_503.epsf}\\(a)}
\parbox{1.5in}{\includegraphics[width=1.4in]{figures/cl/diel/lumo_501_503.epsf}\\(b)}
  \caption{Contour level of HOMOs and LUMOs (equivalent in this case)
    of chlorine interaction with oxygen-terminated $(0001$) surface in
    (a) vacuum and (b) dielectric, side view ($[0001]$ directions runs
    vertically up the page).}  \label{fig:clHOMOLUMO} 
\end{figure}

Turning to the energy levels, Figure~\ref{fig:elevels}c.1 shows the
levels near the gap.  In this case, three two-thirds filled states
(degenerate to within 19~meV$\sim 220$~K) appear just below the top of
the gap, indicating the presence of a hole in the Cr band, which we
interpret as arising from the chlorine atom absorbing an electron from
the chromium oxide to become Cl$^-$. To explore local effects from the
adsorbed chlorine, Figure~\ref{fig:Cldos} presents the local density
of states, weighing each state with the probability of an electron
being within 1.6~\AA\ of the chlorine nucleus.  In contrast to the
local density of states of the hydrogen calculation, the appearance of
the oxygen band is significantly reduced and there is much stronger
mixing with the bulk chromium states.  This mixing corresponds to
alignment of the barely bound Cl$^-$ states with the bulk HOMO
chromium band as the chlorine ion draws electrons from the bulk
chromium band, which is serving as a reservoir of electrons.  Finally,
Figure~\ref{fig:clHOMOLUMO}a shows the sum of the electron
probabilities in the partially filled states at the Fermi level, which
are thus both the HOMOs and the LUMOs.  Interpreted as the sum states
lacking exactly one electron from full occupancy, the figure shows the
spatial distribution of the hole which the formation of the Cl$^-$
injects into the chromium-oxide slab.  As one would expect, this
(positive) hole tends to localize to the vicinity of the Cl$^-$ ions.

Upon relaxation of the physisorbed chlorine surface in the presence of
the solvent, we find there to be very little relaxation (no more than
{0.01 \AA} for any atom), little difference in the total charge
density (Figure~\ref{fig:Cldensvac}b), no change in the presence of
partially filled states at the Fermi level
(Figure~\ref{fig:elevels}c.2), and very little difference in the local
density of states (Figure~\ref{fig:Cldos}) or the spatial distribution
of the hole injected into surface, Figure~\ref{fig:clHOMOLUMO}b.


\subsection{Conclusions}

Above, we introduce the novel approach of using a joint
density-functional theory to treat an {\em ab initio} electronic
structure calculation in the presence of a liquid solvent such as
water.  The resulting approach is the first practical {\em ab initio}
approach for the treatment of the interaction of complex surfaces with
chemical species in the presence of a dielectric environment.

Through this approach, we find the mode of interaction of the
oxygen-terminated Cr$_2$O$_3$ (0001) surface with hydrogen to be
covalent bonding while that with chlorine to be ionic bonding.  The
presence of a dielectric solvent has very little effect on the
pristine surface or on its interaction with chlorine, while it has a
dramatic affect on the interaction with hydrogen.  In vacuum, hydrogen
readily forms an O-H bond with the outermost layer of atoms of the
surface.  In the presence of water, the strong screening
associated with dielectric effects in the vicinity of the proton
(ultimately via hydrogen bonding interactions) so weakens the
attractive potential of the proton that the covalent bond is broken,
the electron is released into the surface and the proton solvates.

In contrast, the interaction with chlorine in vacuum is already ionic,
with a neutral chlorine atom having sufficient electronegativity to
draw an electron from the bulk of the crystal and thus ionize to
Cl$^-$ while injecting a hole into the bulk.  The presence of a
dielectric solvent tends to screen the excess charge on the chlorine
ion, thereby only further stabilizing this form of interaction with
the surface so that there is little change in this case when going
from vacuum to a dielectric environment.  It thus appears that the
primary reason why the solvent has a much greater effect on the
interaction with hydrogen than with chlorine is that a dielectric
environment generally favors the formation of ions and the surface
interaction with chlorine is already ionic whereas the interaction with
hydrogen in vacuum is covalent.

We believe that there would be little effect on these conclusions were
the calculations to be performed with ions rather than atoms.  Doing
so would involve principally removing a single electron from the
calculation for each hydrogen atom or adding an electron for each
chlorine atom.  For the chlorine cases, this would simply remove the
relatively delocalized hole from the bulk chromium band and thus
likely have little effect on the final results.  For the interaction
with hydrogen, the removal of an electron would, in the dielectric
case, simply remove the relatively delocalized donated electron or, in
the vacuum case, likely simply introduce a relatively delocalized hole
into the chromium band.  In either of these cases for hydrogen, we
again would expect little disruption of the chemical integrity of the
surface.

Overall, a novel picture emerges to explain how the oxygen-terminated
surface of Cr$_2$O$_3$ is particularly protective in a hydrochloric
acid solution.  The outer oxygen layer provides a natural barrier to
interaction with chlorine but might be expected to interact strongly
with protons.  However, through dielectric screening effects, it
is the aqueous environment itself which eliminates the outer
oxygen layer's natural tendency to interact with hydrogen.

These first calculations of surface chemistry in the presence of a
solvent make clear the need for additional work to complete the
picture of the passivating effects of chromium oxide.  We would next
like to study the interaction of chlorine and hydrogen with a {\em
chromium}-terminated surface, whose HOMOs would then be exposed on the
surface rather than protected under the outer oxygen layer.  We also
would like to explore pit corrosion by calculating the interaction of
the aforementioned species with a step in the passivating oxygen
layer.  Finally, we would like to contrast the interactions of
chlorine in such systems to the interactions of fluorine and bromine
in order to better understand the corrosive success of chlorine
relative to these other species.

\section{Acknowledgments}
This work was funded by NSF GRANT \#CHE--0113670.  AAR would like to
thank Jefferson W.~Tester and Ronald M.~Latanision for planting the
seed of this project during her 1994--1995 sabbatic leave at MIT.

\bibliographystyle{jpc}

\end{document}